\documentstyle[tighten,preprint,eqsecnum,aps,epsf]{revtex}

\preprint{THU-93/12}

\begin{document}
\draft

\title{
Effects of negative energy components in the constituent quark model}

\author{Peter C. Tiemeijer and J.A. Tjon}

\address{
Institute for Theoretical Physics, University of Utrecht, Box 80.006,
3508 TA Utrecht, The Netherlands.
}

\maketitle

\begin{abstract}
Relativistic covariance requires that in the constituent quark model
for mesons the positive energy states as well as the negative energy
states are included. Using relativistic quasi-potential equations the
contribution of the negative energy states is studied for the light
and charmonium mesons. It is found that these states change the meson
mass spectrum significantly but leave its global structure
untouched.
\end{abstract}

\pacs{PACS numbers: 12.35.H, 12.40.Q, 11.10.Q}

\section{INTRODUCTION}

The application of the constituent quark model to mesons and baryons
has been very successful, in spite of various nonrelativistic
approximations. How much is neglected in these approximations?
Generalization of the ordinary Schr\"odinger equation to a
relativistic covariant form results in the well-known Bethe-Salpeter
equation (BSE), which differs from the nonrelativistic equation in
two respects. Firstly, the dependence of the bound state wave
function on the relative three-momentum $\bbox p$ becomes a
dependence on the four-momentum $(p_0,\bbox p)$. Secondly, covariance
requires that for fermions the full Dirac structure is taken into
account, so for quarks not only the positive energy states are to be
considered, but also their negative energy states must be included.
In this paper we investigate the effect on the meson mass spectrum of
the inclusion of these negative energy states in the relativistic
constituent quark model.

Quasi-potential (QP) approximations to the BSE exist which eliminate
the $p_0$-dependence of the quark-propagators, and moreover
circumvent the problem of introducing a confining potential in the
full Minkowski space. The covariant Dirac structure is kept in these.
In previous work \cite{peter2,peter}, to which we shall refer as I
and II, we used two QP models to calculate the full meson mass
spectrum. One of the biggest differences found with the Schr\"odinger
approach was the much stronger confinement needed and the sensitivity
of the mass spectrum to small vector contributions to the confining
potential. To get some more insight about the various aspects we have
performed calculations in the same framework but with leaving out
various negative energy states. This enables us to trace back and
understand the origin of the phenomena found in I and II and to
establish the importance of the negative energy states. For
simplicity the calculations are restricted to the equal mass case.

In the next section the model is briefly summarized and some of the
QP equations properties which give upper bounds on the parameters of
the potential are discussed. Section III deals with the truncation of
the relativistic QP equations in the coordinate represention with
regards to the negative energy spinor states. It also presents the
comparison of the calculations with corresponding nonrelativistic
predictions. In particular the u\=u and c\=c spectra are considered
in various approximations to various QP models. Differences in the
spectra are discussed. Finally, the paper closes with some concluding
remarks.

\section{REVIEW OF MODEL}

We briefly describe the relativistic constituent quark model as was
studied in I and II. The model consists of two point-like fermions
bound together by a phenomenological potential $V$ to form a meson.
The meson wave function $\psi$ satisfies
\begin{equation}
S^{-1}(\bbox p)\psi(\bbox p)=
- \int \frac{d\bbox q}{(2\pi)^3} V(\bbox p - \bbox q)\psi(\bbox q).
\label{pp1}
\end{equation}
The dependence of the propagator $S$ on the relative four-momentum
$(p_0,\bbox p)$ has been simplified to a dependence on the
three-momentum. In I we studied two QP approximations which eliminate
the dependence of the equation on $p_0$, the
Blankenbecler-Sugar-Logunov-Tavkhelidze (BSLT) approximation
\cite{bslt,cooper}, and an equal-time (ET) approximation
\cite{wallace,wallace2}.
They give for quarks of equal mass in the center of mass system
\begin{equation}
S_{BSLT}^{-1}(\bbox p)=
4\omega\left[
\frac{\omega-E}{\omega+E}\Lambda^{++}
-\Lambda^{+-}-\Lambda^{-+}
+\frac{\omega+E}{\omega-E}\Lambda^{--}\right],
\end{equation}
and
\begin{equation}
S^{-1}_{ET}(\bbox p)=
2(\omega-E)\Lambda^{++}
-2\omega(\Lambda^{+-}+\Lambda^{-+})
+2(\omega+E)\Lambda^{--},
\end{equation}
where $\omega=\sqrt{\bbox p^2+m^2}$, and $E=M/2$, $M$ being the total
meson mass. The $\Lambda^{\rho_1 \rho_2}$ project upon positive and
negative energy states, $\Lambda^{\rho_1
\rho_2}=\Lambda_1^{\rho_1}(\bbox p)
\Lambda_2^{\rho_2}(-\bbox p)$ with
\begin{equation}
\Lambda_i^{\rho_i}(\bbox p) =
\frac{ \rho_i (\bbox \gamma^{(i)} \!\cdot\! \bbox p + m_i)
+\omega_i \gamma_0^{(i)} }{2 \omega_i}.
\end{equation}
Let us define the eigenstates of these projection operators by
\begin{equation}
\Lambda^{\rho_1\rho_2} \gamma_0^{(1)} \gamma_0^{(2)}
|\rho_1 \rho_2\rangle_{pw}
=|\rho_1 \rho_2\rangle_{pw}.
\end{equation}
The subscript $pw$ is used to distinguish these plane wave states
from the canonical positive and negative states defined through
\begin{equation}
\label{can}
\gamma_0^{(1)} \gamma_0^{(2)}
|\rho_1 \rho_2\rangle_{ca}
=\rho_1 \rho_2 |\rho_1 \rho_2\rangle_{ca},
\end{equation}
which correspond to the various combinations of upper and lower
components of the Dirac spinors. Clearly
\begin{equation}
|\rho_1 \rho_2\rangle_{pw}
=|\rho_1 \rho_2\rangle_{ca}+ {\cal O}(p/m).
\end{equation}

The instantaneous interaction $V$ between the quarks is modeled as
the sum of a Coulomb-like part describing the one-gluon-exchange
(OGE) interaction and a linearly rising part for the confinement. It
takes in coordinate space the form
\begin{equation}
V(x)=-\frac{\alpha(x)}{x} \Gamma_V
+(\kappa x + c) \left[ (1-\varepsilon) 1^{(1)} 1^{(2)}
+ \varepsilon \Gamma_V \right].
\label{pp5}
\end{equation}
The vector contribution to the interaction is studied in the Feynman
gauge as well as in the Coulomb gauge:
\begin{eqnarray}
\Gamma_V^{Feynman}&=& \gamma_\mu^{(1)} \gamma^{\mu (2)},
\label{pp8}
\\
\Gamma_V^{Coulomb}&=& \gamma_\mu^{(1)} \gamma^{\mu (2)}
+\frac{1}{2} \left[
\bbox\gamma^{(1)} \!\cdot\!\bbox \gamma^{(2)}
-(\bbox \gamma^{(1)} \!\cdot\! \hat{\bbox x})
(\bbox \gamma^{(2)} \!\cdot\! \hat{\bbox x})
\right].
\end{eqnarray}
This defines the relativistic quasi-potential model. Note that we do
allow for a fraction $\varepsilon$ of vector-confinement. The running
coupling constant behaves as $\alpha(x) \sim (8/27)\pi/\ln(x_0/x)$
for small distances $x$, and grows to some maximum saturation value
$\alpha_{sat}$ for large separations, according to the interpolation
given in I.

The resulting wave equation were studied extensively in coordinate
space. It should be noted that certain difficulties as found in the
one particle Dirac equation should be expected in such a relativistic
quasi-potential approach. Let us for a moment consider one fermion
in an external potential. If this particle experiences a potential
which fluctuates more strongly than $\sim 2m$ over a distance shorter
than its Compton length $x_C=1/m$ new fermion-antifermion pairs can
be created. This phenomenon cannot correctly be described by the
Dirac equation which describes a one-particle theory and thus misses
the interactions between the newly created pair and the starting
particle. Since the Dirac equation does allow for antifermion
components, solutions in this potential can have an unbound number of
non-interacting fermions and antifermions thus being unnormalizable
and unphysical. This break-down of the Dirac equation is well-known
as the Klein paradox \cite{bd}. Similar flaws emerge in the QP
equations of this work since they also contain negative energy
components. In view of the complexity of the various two-body
equations we do not fully analyze under what conditions they break
down. Instead, let us note that unbound solutions can be expected if
the confining strength becomes too strong, $\kappa x_C \gtrsim 2m$ or
$\kappa \gtrsim 2m^2$. This domain can be reached in light meson
systems. The condition on $\kappa$ depends on the fraction
$\varepsilon$ of vector confinement as indicated by the discussion in
I. Similarly, if the OGE potential becomes too strong, $\alpha/x_C
\gtrsim 2m$ or $\alpha \gtrsim 2$, irregular solutions may be
expected. Mesons with high orbital angular momenta are less sensitive
to this effect because the centrifugal barrier prevents them from
entering the short distance region. If a running coupling constant is
taken instead of a fixed one then the irregular solutions disappear
and the singular behavior becomes less. Whereas the upper bound on
$\kappa$ disappears as the negative energy components are removed,
the upper bound on $\alpha$ is shifted upward but still present,
reflecting that only positive energy states can also tumble in a OGE
potential. Detailed discussions on the short distance behavior was
given in I and II.

\section{RESULTS}

Since the wave equations as described in the previous section are
solved in the coordinate space using the representation of the
canonical states Eq.~(\ref{can}), a convenient way to switch off the
coupling to the negative energy plane wave states is to explicitly
project out these states in the interaction. The projection is done
by rewriting the BSLT equation as
\begin{eqnarray}
\lefteqn{ \left[
(\omega-E)\Lambda^{++}
-E(\Lambda^{+-}+\Lambda^{-+})
-(i_{\flat\flat}\omega+E)\Lambda^{--}
\right] \psi =}\nonumber\\
&&
-\frac{1}{4\omega}\left[ (\omega+E)\Lambda^{++}
+i_\flat E(\Lambda^{+-}+\Lambda^{-+})
+i_{\flat\flat}(E-\omega)\Lambda^{--}
\right] \gamma_0^{(1)} \gamma_0^{(2)} V \psi.
\label{pp2}
\end{eqnarray}
For $i_\flat=i_{\flat\flat}=1$ we recover the QP equations with all
negative energy states included. If $i_\flat$ is put to zero then the
single negative energy states are projected to zero, provided $E\ne
0$, and similarly for $i_{\flat\flat}$ when we want to drop the
coupling to the $|\!--\rangle_{pw}$ states in the calculations. One
can rewrite the ET equation in a similar way:
\begin{eqnarray}
\lefteqn{ 2\left[
(\omega-E)\Lambda^{++}
-{\omega}(\Lambda^{+-}+\Lambda^{-+})
+(i_{\flat\flat}\omega+E)\Lambda^{--}
\right] \psi =}\nonumber\\
&&
-\left[
\Lambda^{++}
+i_\flat(\Lambda^{+-}+\Lambda^{-+})
+i_{\flat\flat}\Lambda^{--}
\right] \gamma_0^{(1)} \gamma_0^{(2)} V \psi.
\label{pp3}
\end{eqnarray}

Although the number of coupled channels do not change, the above
Eqs.~(\ref{pp2}) and (\ref{pp3}) have a major advantage that they
can be solved in the same way as the full BSLT and ET equations were
solved in I. First we Fourier transform them to configuration space
and make an angular momentum decomposition. This gives a set of
coupled integral-differential equations which are reduced to a set of
linear equations by expanding the wave function on a set of spline
functions. The resulting matrix equation for the various spline
coefficients can be solved straightforwardly by standard methods. As
a check on the accuracy of our calculational procedures we used this
method to calculate the charmonium mass spectrum in the model of
Hirano
{\em et$\;$al.} \cite{murota}, which includes only the
$|\!+\!+\rangle_{pw}$ and $|\!--\rangle_{pw}$ states. Agreement was
found within 1 MeV.

\subsection{Comparison with Schr\"odinger equation results}
\noindent
To study the effects of various relativistic contributions we
consider the mass spectra for the light mesons and the charmonium
system. In the limit of large quark masses the BSLT can be reduced to
the nonrelativistic Schr\"odinger equation
\begin{equation}
\left[
2\left(-\frac{\bbox \nabla^2}{2m}+m \right)+V_{NR}(x)+V_{SD}(x)
\right] \psi(x)
= M \psi(x),
\label{pp4}
\end{equation}
where $V_{NR}$ is the nonrelativistic reduction of the potential
\begin{equation}
V_{NR}(x)=-\frac{\alpha(x)}{x}+\kappa x +c,
\end{equation}
and $V_{SD}$ contains the spin-dependent corrections of order $1/m^2$
\cite{lucha}
\begin{eqnarray}
\label{nrpot}
V_{SD}(x)& =&\frac{1}{m^2} \left[
\frac{3V'_V-V'_S}{2x}
\bbox L \!\cdot\! \bbox S
+\frac{2}{3} (\bbox \nabla^2 V_V) \bbox S_1 \!\cdot\! \bbox S_2
+\left( \frac{1}{x}V'_V - V''_V \right)S_{12} \right],
\nonumber \\
S_{12} & = & (\bbox S_1 \!\cdot\! \hat{\bbox x}) (\bbox S_2 \!\cdot\!
\hat{\bbox x})
-\frac{1}{3} \bbox S_1 \!\cdot\! \bbox S_2.
\end{eqnarray}
Here $V_V$ and $V_S$ denote the vector and scalar contributions to
the potential. The potential in Eq.~(\ref{nrpot}) is singular at the
origin and therefore a regularization is needed. The delta function
appearing in $\bbox \nabla^2 V_V$ is replaced by a gaussian of width
$\beta^{-1}_1$, and the $x^{-1}$- and $x^{-3}$-singularities are cut
off at $x_0=\beta_2^{-1}$ and replaced by smoothly fitting gaussians.
We take the same parameters for the potential as in I, which are
summarized in table 1, and take the $\beta_i$ such that they
reproduce approximately the same splittings between the $S$- and
$P$-states as the full BSLT calculation. With these parameters we
find the spectra as shown in the first column of Figs.~1 and 2 of
the u\=u and c\=c systems. The spectra of the various columns in
these figures will be referred to as Fig.~1a, 1b, ... etc.

The first natural step in relativizing the Schr\"odinger equation is
to replace the kinetic energy according to
\begin{equation}
\frac{\bbox p^2}{2m}+m \rightarrow \omega=\sqrt{\bbox p^2+m^2}
\end{equation}
in Eq.~(\ref{pp4}). Figs.~1b and 2b show the corresponding spectra.
They are shifted downwards as compared to the nonrelativistic ones
indicating that the interaction has become more attractive. Also the
shift becomes larger as the level of excitation increases. For
charmonium one finds shifts between 0.05 -- 0.13 GeV, whereas for the
light mesons these shifts are more substantial and range between 0.5
-- 1.5 GeV.

The second relativistic correction which can be included is to
replace the approximation of the potential $V_{NR}+V_{SD}$ by the
complete projection of the potential Eq.~(\ref{pp5}) on positive
energy states $V^{+\!+,+\!+} \equiv\ _{pw}\langle +\!+|V
\gamma_0^{(1)} \gamma_0^{(2)}
|+\!+\rangle_{pw}$. The singular behavior of $V_{SD}$ is no longer
present in the complete projection, and hence the cut-off parameters
$\beta_i$ are principly absent in the resulting wave equations. Note
the factors $\gamma_0^{(1)} \gamma_0^{(2)}$ in $V^{+\!+,+\!+}$ and in
the right hand sides of Eqs.~(\ref{pp2}) and (\ref{pp3}) which
reflect that the energies of the quarks are the fourth components of
their four-momenta. This replacement gives a rather large effect
(0.00 -- 0.16 GeV for c\=c and 0.0 -- 0.8 GeV for u\=u) as is
illustrated by the spectra in Fig.~1c and 2c and which is easily
understood. For a potential $V(x)=V_S(x) \bbox I + V_V(x)
\gamma_0^{(1)} \gamma_0^{(2)}$ and low momenta of the in- and
outgoing states
\begin{equation}
V^{+\!+,+\!+} \approx
\ _{ca}\langle +\!+|V_S \gamma_0^{(1)}\gamma_0^{(2)}+V_V(x) \bbox I
|+\!+\rangle_{ca}
= V_S(x)+V_V(x).
\end{equation}
But for relativistic momenta considerable contributions to
$V^{+\!+,+\!+}$ are to be expected from the contributions
\begin{equation}
_{ca}\langle -\!+|V_S \gamma_0^{(1)}\gamma_0^{(2)}+V_V(x) \bbox I
|-\!+\rangle_{ca}
= -V_S(x)+V_V(x).
\label{pp6}
\end{equation}
For scalar confinement it leads to a considerable reduction of the
confinement strength. This effect can also qualitatively be seen from
the spin-independent corrections of order $1/m^2$ \cite{lucha} to a
nonrelativistic scalar potential. It is given by
\begin{equation}
V_{SI}(x)=\frac{1}{m^2}\left[
\frac{1}{4}
(\bbox \nabla^2 V_S)+V'_S \frac{d\hfill}{dx}+V_S \bbox \nabla^2
\right].
\end{equation}
This potential, however, does not give an accurate approximation
since for large distances $m^{-2}V_S\bbox \nabla^2$ is not a small
enough parameter to expand in.

To obtain the explicit form of the BSLT propagator we may replace
\begin{equation}
(\omega-E) \rightarrow \frac{2\omega}{\omega+E}(\omega-E).
\label{pp7}
\end{equation}
This leads to the BSLT equation restricted to only positive energy
states. Figs.~1d and 2d show how the masses are increased.

Figure 1e shows the effect of the introduction of the single negative
energy states $|\!+\!-\rangle_{pw}$ and $|\!-\!+\rangle_{pw}$. For
the light mesons we do not show this case since for some mesons it
leads to unbound systems, similar to the Klein paradox mentioned in
the previous section. The introduction of these states leads to an
increase of the c\=c spectrum by up to 0.10 GeV. They are mostly made
out of $|\!+\!-\rangle_{ca}$ and $|\!-\!+\rangle_{ca}$ states which
have a negative expectation value of scalar confinements,
Eq.~(\ref{pp6}). Together with the negative propagation of these
states this leads to a positive interaction which raises the mass
levels.

Finally, if the $|\!--\rangle_{pw}$ states are included, one arrives
at the full BSLT spectra shown in Figs.~1f and 2e. The inclusion of
these states changes little ($\lesssim$ 1 MeV for c\=c), as could be
expected from their smallness. The lowest lying mesons, however, are
considerably influenced what is related to the singular behavior of
the wave function at small distances. The OGE potential concentrates
the wave function to small distances with strong potential, thus
lowering the mass levels. The singular behavior becomes stronger if
more negative energy states are included or if the level of
excitation is less. Fig.~1g shows the charmonium spectrum in the
Salpeter approximation, which includes only the $|\!+\!+\rangle_{pw}$
and $|\!--\rangle_{pw}$ states. The small masses for the low lying
excitations are still present, but the increase associated with the
single negative states is absent.

\subsection{Equal-time results}
\noindent
Let us briefly discuss what effects are found if we use the equal
time (ET) propagator. In Fig.~3 is shown the corresponding charmonium
spectrum calculated with the same parameters as Fig.~1. Three
differences can be noted. Firstly, the ET spectrum is lowered as
compared to the BSLT one. This is a consequence of the different
propagators for the $|\!+\!+\rangle_{pw}$ states
[cf.\ Eq.~(\ref{pp7})]. Secondly, the introduction of the single
negative energy states causes a shift upwards in the ET model which
is approximately twice that seen in BSLT. This is due to the fact
that these states are roughly twice as important in the ET wave
function as in the BSLT one in view of a twice as large propagator
for these states, $S_{ET}^{++} = -1/2\omega$ versus $S_{BSLT}^{++} =
-1/4\omega$. Thirdly, the $|\!--\rangle_{pw}$ energy states lower the
singlet states much more strongly in the ET model than in BSLT.
Again, this is a consequence of the importance of these states in the
ET wave function. They are larger due to the large $S_{ET}^{--}$
propagator. As already remarked in \cite{wallace2}, this is a defect
of the ET equation when an unretarded potential is used, and this
disappears if the proper retardation is inserted in the matrix
elements for $V$ that connect to double negative energy states.
However, it should be stressed that there is no unambigious extension
of the instantaneous confining potential to retarded times.

\subsection{Sensitivity to vector confinement}
\noindent
In I we discussed the large raising of the light meson spectrum when
a part of vector confinement is added to a scalar confining
potential. Fig.~4 shows the u\=u spectrum for various dynamical
models using scalar confinement without and with a vector
contribution. When only positive energy states are taken the raising
is even more stronger, and can almost double some meson masses.

In Fig.~4 are also shown the same spectrum if the vector contribution
is taken in the Feynman gauge, Eq.~(\ref{pp8}). In this gauge the
spin-spin interaction is less suppressed so the vector contribution
has more effect.

\section{Concluding remarks}

In summary, we have calculated for the light and charmonium systems
mass spectra using the (relativized) Schr\"odinger equation and the
relativistic quasi-potential equations of I and II, which differ from
the Schr\"odinger approach in that no $1/m^2$ approximation is made,
and that the QP equations contain the full Dirac structure of
positive and negative energy states. We studied the importance of
these differences by solving the QP equations while leaving out
negative energy components. We find that the projection of the
confining potential on positive energy states leads to a considerably
lower confinement strength than the nonrelativistic potential gives.
This is partly compensated for by the introduction of the single
negative energy states which are more bound than the positive states,
and hence increase the masses. Finally we find that the double
negative energy states have little influence except for short
distances where the singular behavior of the mesons is strengthened.

The total picture stongly suggests that the differences here studied
are important for determining the parameters of the q\=q interaction,
especially the confinement strength $\kappa$. Yet the global
structure of the spectrum ---level ordening, relative sizes of
splittings--- remains rather untouched under the relativistic
modifications. This confirms the conclusion drawn from the success of
nonrelativistic quark models that most relativistic effects in q\=q
spectroscopy can be mimicked by employing nonrelativistic dynamics
together with effective parameters. It will be interesting to study
whether the relativistic modifications to meson wave functions will
lead to sizeable changes in cross section for processes such as e.m.
ones involving these mesons.

In this paper we have studied relativistic effects within the
framework of quasi-potential equations. In so doing we have not
addressed entirely the role of the relative energy variable $p_0$.
Apart from the complexity of a calculation including this say in a
Bethe-Salpeter equation approach, a more fundamental obstacle is
posed by the extension of the definition of the confining potential,
as we have used here, to a four-momentum dependence. The confining
potential is only known for the static case. There is at this moment
no underlying theory which can give a prescription on how to extend
it to a covariant form. Let us illustrate this considering the
commonly used generalization of the potential $V(x)=\kappa x$ which
reads in momentum space
\begin{equation}
V(q_0,\bbox q)=\frac{\kappa}{2\pi^2}
\lim_{\eta \rightarrow 0}
\left[
\frac{\partial^2\hfill}{\partial \eta^2}
\frac{1}{\bbox q^2-q_0^2+\eta^2(1-i\varepsilon)}
\right],
\end{equation}
but its Fourier transform yields
\begin{equation}
V(t,x)=\frac{\kappa}{\pi}
\lim_{\eta \rightarrow 0}
\left[
K_0(\eta R)- \eta R K_1(\eta R)
\right]
= \infty,
\end{equation}
(with $R^2=x^2-t^2$) which is clearly physically unacceptable.
Exploring QCD may lead to ways of reconstructing such a confining
force, which can be used in such an off mass shell approach as we
have discussed here.

\acknowledgments
This work was partially financially supported by de Stichting voor
Fundamenteel Onderzoek der Materie (FOM), which is sponsored by the
Nederlandse Organisatie voor Wetenschappelijk Onderzoek (NWO).

\begin{figure}
\epsfxsize=\textwidth
\epsfbox{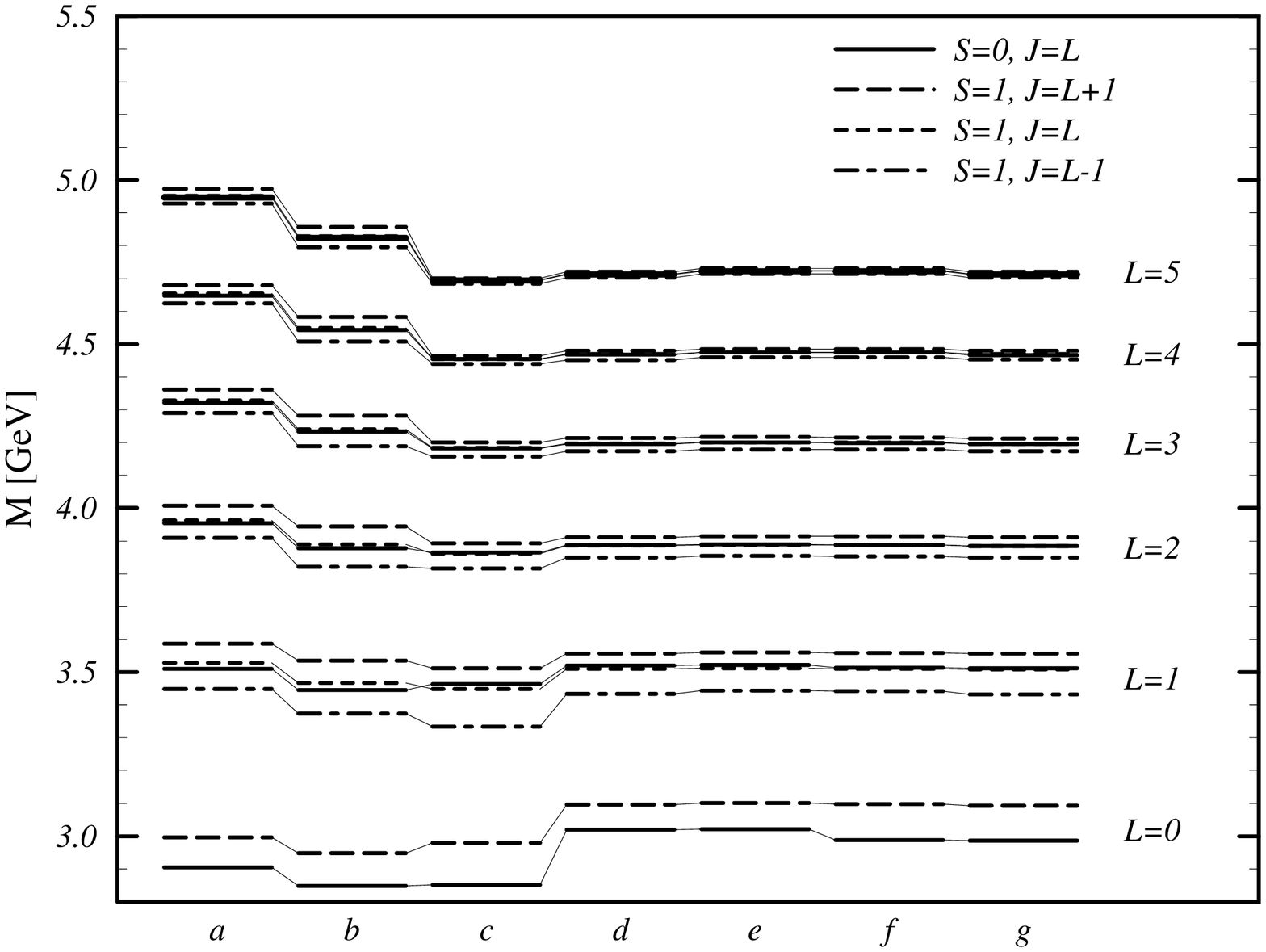}
\caption[fig31]{
Charmonium spectrum of the radially unexcited states.
\bbox a:~Schr\"odinger equation with $p^2/2m$ and Breit interaction,
\bbox b:~Schr\"odinger equation with $\sqrt{p^2+m^2}$ and
Breit interaction,
\bbox c:~Schr\"odinger equation with $\sqrt{p^2+m^2}$ and
full projection of potential in Coulomb-gauge, i.e. ET with
$|+\!+\rangle_{pw}$,
\bbox d:~BSLT with $|+\!+\rangle_{pw}$,
\bbox e:~BSLT with $|+\!+\rangle_{pw}$,
$|\!+\!-\rangle_{pw}$ and $|\!-\!+\rangle_{pw}$,
\bbox f:~BSLT with all states,
\bbox g:~BSLT with $|+\!+\rangle_{pw}$ and $|\!--\rangle_{pw}$.}
\label{fig31}
\end{figure}

\begin{figure}
\epsfxsize=\textwidth
\epsfbox{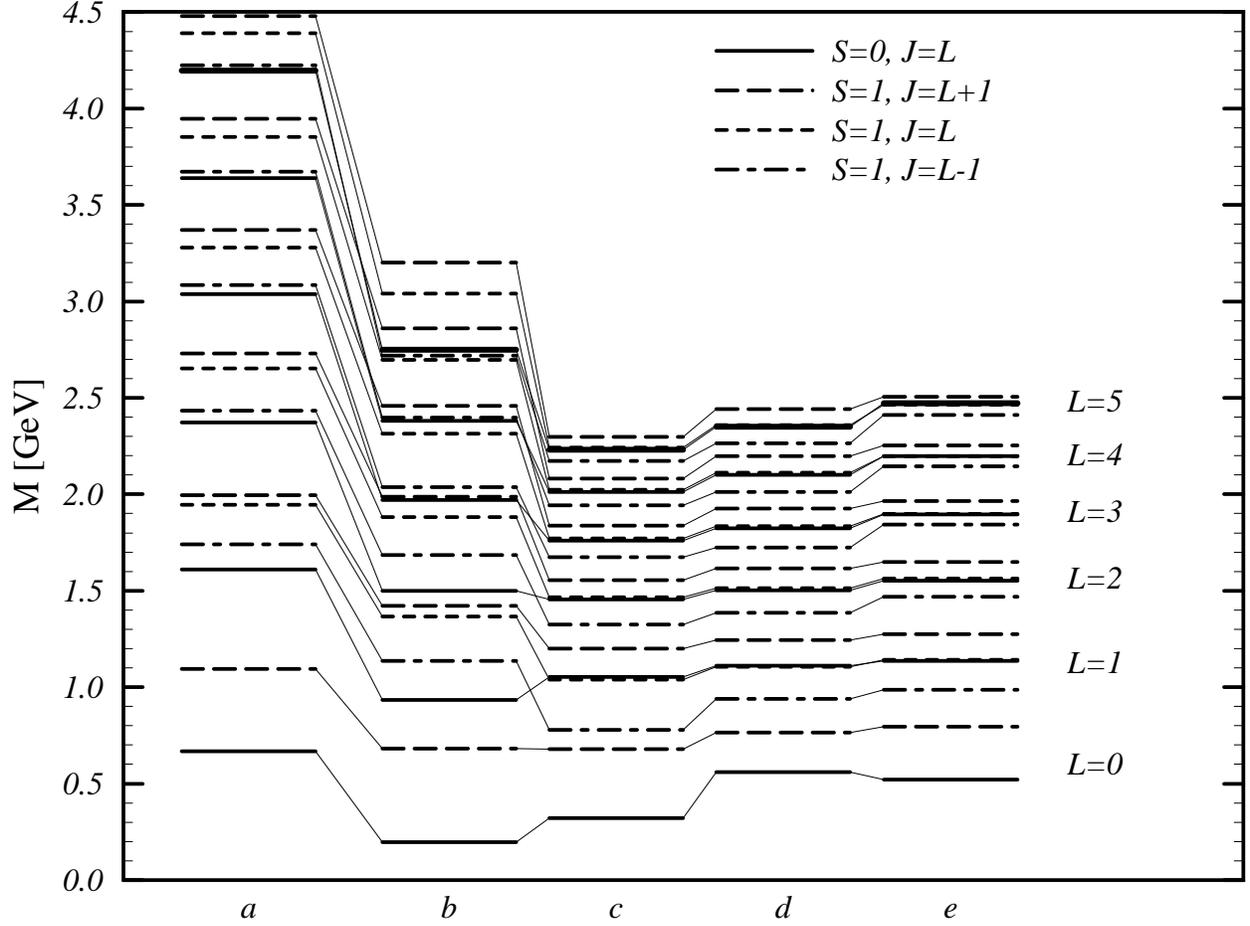}
\caption[fig32]{
Light meson spectrum of the radially unexcited states.
\bbox a:~Schr\"odinger equation with $p^2/2m$ and Breit interaction,
\bbox b:~Schr\"odinger equation with $\sqrt{p^2+m^2}$ and
Breit interaction,
\bbox c:~Schr\"odinger equation with $\sqrt{p^2+m^2}$ and
full projection of potential in Coulomb-gauge, i.e. ET with
$|+\!+\rangle_{pw}$,
\bbox d:~BSLT with $|+\!+\rangle_{pw}$,
\bbox e:~BSLT with all states.}
\label{fig32}
\end{figure}

\begin{figure}
\epsfxsize=\textwidth
\epsfbox{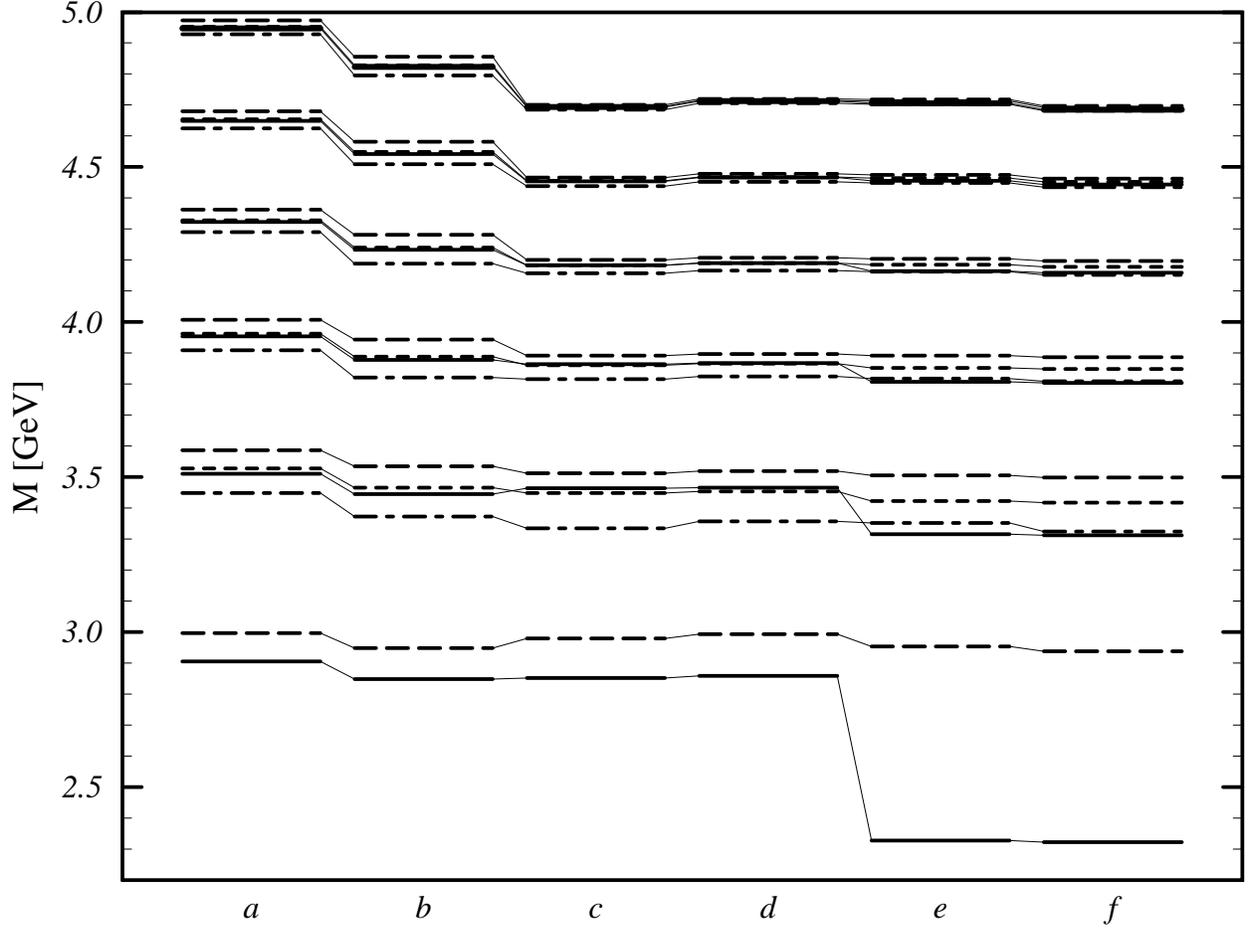}
\caption[fig33]{
Charmonium spectrum of the radially unexcited states (legend as in
Fig.~\ref{fig31}).
\bbox a:~Schr\"odinger equation with $p^2/2m$ and Breit interaction,
\bbox b:~Schr\"odinger equation with $\sqrt{p^2+m^2}$ and
Breit interaction,
\bbox c:~Schr\"odinger equation with $\sqrt{p^2+m^2}$ and full
projection of potential, i.e. ET with $|+\!+\rangle_{pw}$,
\bbox d:~ET with $|+\!+\rangle_{pw}$,
$|\!+\!-\rangle_{pw}$ and $|\!-\!+\rangle_{pw}$,
\bbox e:~ET with all states,
\bbox f:~ET with $|+\!+\rangle_{pw}$ and $|\!--\rangle_{pw}$.}
\label{fig33}
\end{figure}

\begin{figure}
\epsfxsize=\textwidth
\epsfbox{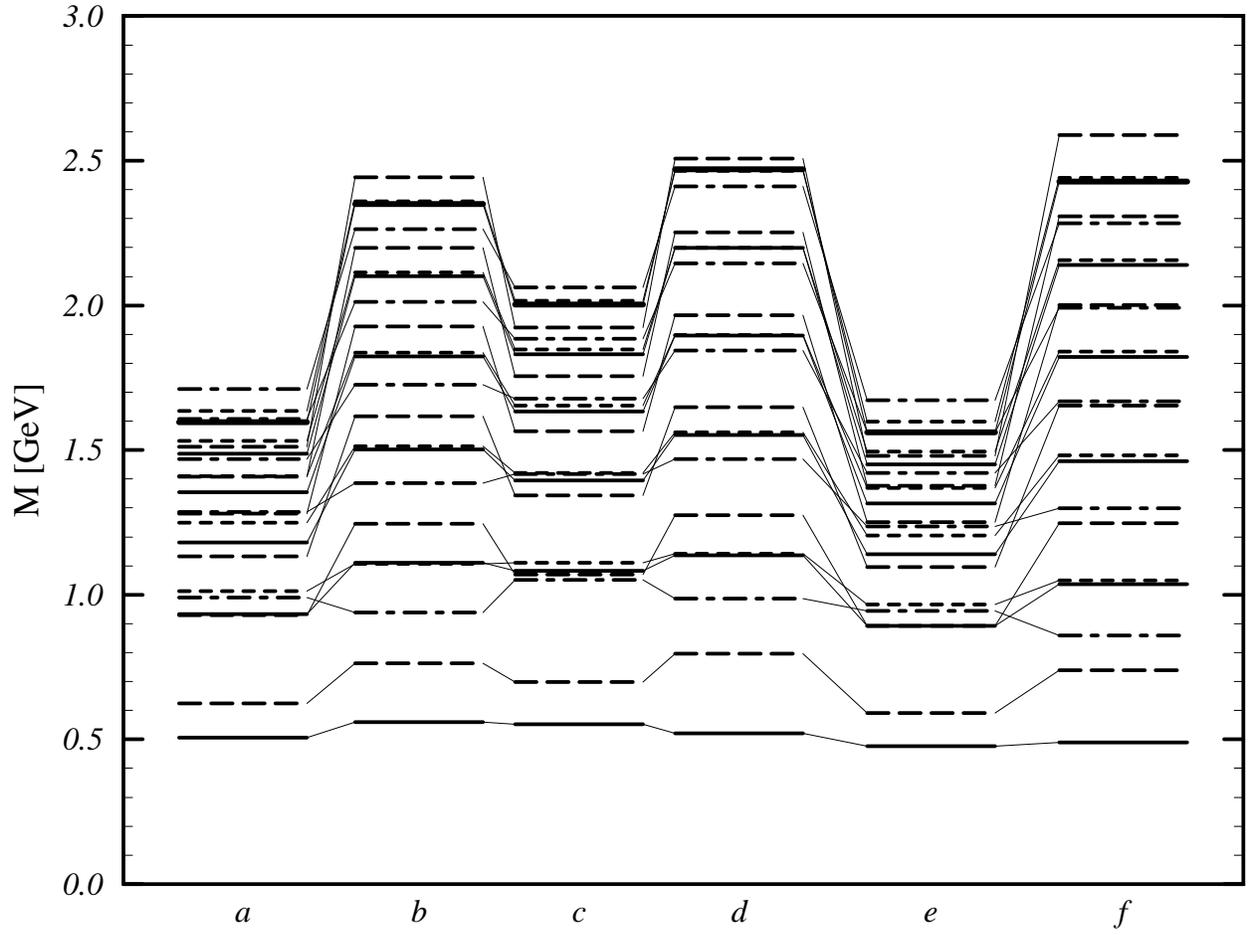}
\caption[fig34]{
Light meson spectrum of the radially unexcited states (legend as in
Fig.~\ref{fig31}).
\bbox a:~BSLT in Coulomb gauge with $|+\!+\rangle_{pw}$ and
$\varepsilon=0$,
\bbox b:~BSLT in Coulomb gauge with $|+\!+\rangle_{pw}$
and $\varepsilon=0.25$,
\bbox c:~BSLT in Coulomb gauge with all states and $\varepsilon=0$,
\bbox d:~BSLT in Coulomb gauge with all states and
$\varepsilon=0.25$,
\bbox e:~BSLT in Feynman gauge with $|+\!+\rangle_{pw}$ and
$\varepsilon=0$,
\bbox f:~BSLT in Feynman gauge with $|+\!+\rangle_{pw}$
and $\varepsilon=0.25$.}
\label{fig34}
\end{figure}

\mediumtext

\begin{table}
\caption{Parameters for the q\=q model}
\begin{tabular}{lddl}
& u\=u & c\=c &\\
\tableline
$m$               & 0.250 & 1.779 &GeV\\ $\kappa$          & 0.33  &
0.33  &GeV$^2$\\ $c$               &-1.0   &-1.0   &GeV\\
$\varepsilon$     &0.25   & 0.25  &\\ $\alpha_{sat}$\tablenotemark[1]
		  &0.8    & 0.8  & \\ $\beta_1$         & 0.25  & 0.8
&GeV\rule{0ex}{4ex}\\ $\beta_2$         & 0.10  & 0.6  &GeV\\
\end{tabular}
\label{table31}
\tablenotetext[1]
{Running as $\alpha_I$ in Ref.~\cite{peter2} with $\mu=1.0$}
\end{table}

\end{document}